# Accessible Interactive Maps for Visually Impaired Users

*Julie Ducasse[1,2]\*, Anke Brock[3]\*, Christophe Jouffrais[1,2]*

\* Both authors are 1[st] co-authors and contributed equally to the chapter

1 : CNRS, IRIT, Toulouse, France
2 : University Paul Sabatier, IRIT, Toulouse, France
3 : INRIA, Labri, Bordeaux, France

## Contenu

# I. Introduction

## I.1 Context

Mobility and orientation are among the greatest challenges for visually impaired people. In France, 56% of the visually impaired population admit to facing difficulties when walking outside and 29% are not able to navigate on their own (C2RP, 2005). One reason that explains these issues is that visually impaired users usually exchange verbal descriptions of itineraries, which may help them to find their way, but do not provide them with any clue about the spatial layout of the targeted environment. GPS-based systems, although facilitating navigation, raise the same issue. Sighted people usually acquire information about a spatial environment through visual perception or by using geographical maps. However, maps are essentially visual and thus inherently inaccessible for visually impaired people. And weak access to maps has drastic consequences on mobility and education, but more generally on personal and professional life, and can lead to social exclusion (Passini & Proulx, 1988).

Beyond orientation and mobility purposes, maps are very useful tools to explore and analyze geographical data and to acquire general knowledge about many subjects such as demography, geopolitics, history, etc. They are also often used in the classroom for this purpose. As stated by O'Modhrain *et al.* (2015), "there is an immediate need for research and development of new technologies to provide non-visual access to graphical material. While the importance of this access is obvious in many educational, vocational, and social contexts for visually impaired people, the diversity of the user group, range of available technologies, and breadth of tasks to be supported complicate the research and development process".

In specialized educational centers for visually impaired people, tactile maps are commonly used to give visually impaired students access to geographical representations. However, these materials are rarely used or available outside of school, because their production is a costly and time-consuming process (Rice, Jacobson, Golledge, & Jones, 2005). To create a tactile map, one of the most common methods is to print it on a special paper, called "swell paper" (synonyms are "microcapsule paper" or "heat sensitive paper"), which contains microcapsules of alcohol in its coating. When the paper is heated, the microcapsules expand and create relief over black lines (Figure 1.a). The resulting maps, called raised-line maps, can be perceived by touch. But they are also visual maps, making it possible to share information between blind and (partially) sighted people. Raised-line maps are usually prepared with a computer, which makes it possible to print and fuse several copies of the same map.

Another techniques, called vacuum-forming or thermoforming, consists in placing a sheet of plastic over a master made of a variety of textured materials. When it is heated in a vacuum, the sheet is permanently deformed according to the master. Hand-crafted techniques can also be used to produce maps and other graphics. For example, for orientation and mobility lessons, teachers and students construct itineraries or local maps, by progressively placing magnets over a magnetic board (Figure 1.b). Students are sometimes asked to replicate the construction, so that the teacher can check that the itinerary has been memorized. Small-scale models made out of wood also exist, alongside graphics made out of paper, cardboard, ropes, etc. (see Figure 1.c). Edman (1992) presented a comprehensive summary of production techniques for accessible maps.

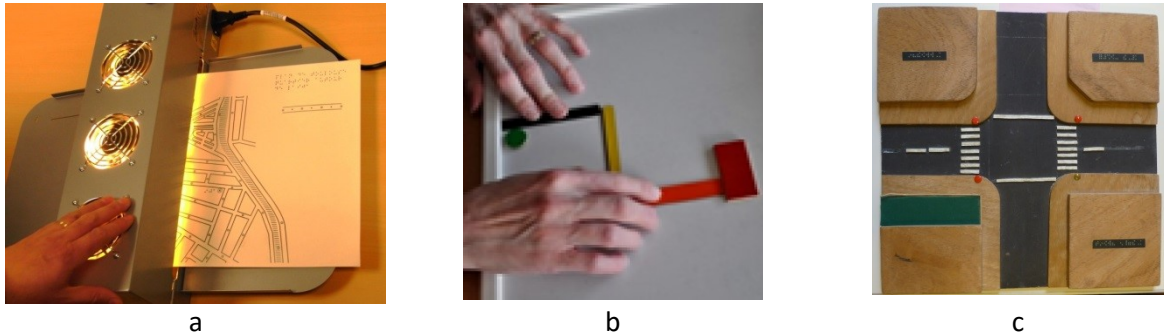

Figure 1: a) Production of a raised-line map. The map is printed on swell paper that is passed into a fuser. The relief appears over black lines. b) A simple local map is being built with small magnets. c) Example of a wooden small-scale model used during Orientation & Mobility lessons.

Although tactile maps are efficient for acquisition of spatial knowledge, they present several limitations and issues. As stated by Rice *et al.* (2005) the production of tactile maps is time consuming and expensive. In addition, tactile maps must be produced by a tactile graphics specialist who knows how to present information so that it can be perceived by touch (Tatham, 1991). Other critiques concern the number of elements that can be displayed and the accuracy of the map content. Indeed, because of the perceptual limits of the tactile modality, less details can be represented on a tactile map than on a visual map. Furthermore, once a tactile map is printed, its content is static and cannot be adapted dynamically. Tactile maps are then quickly getting outdated (Yatani, Banovic, & Truong, 2012). In addition, the use of braille labels is an issue. Only a small percentage of visually impaired people read braille (National Federation of the Blind, 2009); and braille is not so convenient when compared to printed text. Text on visual maps can be written with different font sizes and styles. It can be rotated to fit in open spaces. Upper and lower cases, as well as color, can be used to highlight important items. In contrast, braille lacks all of these possibilities and needs a lot of space because it is fixed in size, inter-cell spacing and orientation (A. F. Tatham, 1991). Due to the lack of space, braille abbreviations are commonly used on tactile maps, which are then explained in a legend accompanying the map. Because the reader must frequently move hands between the map and the legend, it disrupts the reading process (Hinton, 1993). Finally, because they are not interactive, tactile maps cannot provide advanced functionalities such as panning, zooming, annotating, computing itineraries or distances, and using filters to facilitate the exploration.

Several research projects have been led during the past three decades in order to improve access to maps using interactive technologies. In this chapter, we first present an exhaustive review of interactive map prototypes. We classified existing interactive maps into two categories: *Digital Interactive Maps (DIMs)* that are displayed on a flat surface such as a screen; and *Hybrid Interactive Maps (HIMs)* that include both a digital and a physical display. In each family, we identified several subcategories depending on the technology being used. We compared the categories and subcategories according to cost, usability, and functionality. Then, because evaluation is essential in order to verify that devices are usable, we reviewed a number of studies showing that they can support spatial learning for visually impaired users. In the following section, we discussed the similarities and differences between maps and other types of graphics (such as bar charts, organigrams, etc.) Finally, we identified new technologies and methods that could improve the accessibility of graphics for visually impaired users (e.g. shape-changing interfaces and 3D printing).

## I.2 Important notions

### I.2.1 About Accessible Interactive Maps

*Definition and characteristics of accessible interactive maps*

In this chapter, we use the term "map" as suggested by Montello (1993). Montello defines four psychological spaces (i.e. space as perceived by a human) that depend on "the projected size of the space relative to the human body". *Figural space* is "projectively smaller than the body" and can be apprehended without moving. Within figural space, *pictorial spaces* are small flat spaces (i.e. maps) and *object spaces* are 3D spaces (i.e. small scale models). *Vista space* is "as large as or larger than the body" and can be apprehended without moving (e.g. town squares). *Environmental space* is "projectively larger than the body" and requires locomotion to be apprehended (e.g. cities). *Geographical space* is "much larger than the body" and cannot be apprehended through locomotion (e.g. countries). Maps are then pictorial spaces that represent environmental or geographical spaces.

Maps can have different purposes. Reference maps support orientation and mobility, by providing for example the possibilities to explore unknown areas, acquire an overview of the surrounding of a landmark, localize specific landmarks, or prepare a journey. You-are-here (YAH) maps are reference maps showing features in the immediate environment of the map reader (Montello, 2010). They normally include a symbol representing own location, and possibly the orientation of a person viewing the map.

Other map types include topological maps showing selected features without necessarily respecting distance, scale and orientation, as well as thematic maps giving qualitative or quantitative information on a specific topic (Edman, 1992). Cloropleth maps are a subset of thematic maps in which the shading of areas indicates the value of a specific variable (Zhao, Plaisant, Shneiderman, & Lazar, 2008). Map representations are not perfect representations of the environment. Deformations occur when the cartographer transforms spatial data into a map representation. The different map types can be used for teaching geography, but also politics, economy, history, etc.

These observations apply to maps made for both sighted and visually impaired users. However, even though purposes and motivations for using maps are similar for both groups, the map representations are different. They are essentially visual when designed for sighted users, whereas they are based on auditory and tactile cues when designed for visually impaired users.

*Map data vs. map display*

We distinguish between map data and map representation. The map data is usually stored in a database and cannot be explored as it is: it consists in a set of locations (cities, landmarks, streets, etc.), each being determined in absolute terms—latitude, longitude—or in relative terms—in comparison to a reference point (Lloyd, 2000). The map data needs to be rendered in a perceptible way, with visual, haptic, tactile or auditory cues.

When the map data is represented in the form of objects that one can explore by touch, we consider that the map display is physical (e.g., raised-line maps, 3D-printed maps, maps built using magnets, etc.). If such a physical display is also associated with a digital representation, we consider that the map display is hybrid (e.g., an interactive tactile map). When there is no physical display associated with the map, we considered it as digital (e.g. an auditory map explored with the finger).

### *Accessible Interactive Maps*

In this chapter, the term "interactive maps" refers to all the devices that have been designed to display *pictorial spaces*, as defined by Montello (1993). The interaction relies on technologies that provide feedback in response to user's input. Accessible Interactive Maps are designed for visually impaired users. The family of "touch-enabled" devices has been widely used to design Accessible Interactive Maps. They refer to different technologies that directly sense the user's touch input. They have also been called touchscreens or touch-sensitive devices. Some of these devices are mono-touch (e.g. digitizer tablets, older touch-screen models), but most of the current ones are multi-touch (e.g. tablets and smartphones). Some devices react to bare fingers, whereas others require a pen for input (e.g., digitizer tablets). These devices vary in scale, from smartwatches to large tabletops.

### I.2.2  About touch

#### *Sense of touch*

The terms "touch", "haptic", "tactile", etc. are used in many scientific fields including Human-Computer Interaction and Psychology, but with slightly different meanings. Here, we more likely refer to the literature in psychology for their definition.

The sense of touch relies on the somatosensory system which includes thermoreceptors, mechanoreceptors, chemoreceptors and nociceptors. Tactile (or sometimes touch) perception concerns the contact between the skin of any part of the body and other objects. Because of the presence of different receptors in the skin, muscles, and bones, the sense of touch provides cues on temperature, pressure, movement, and pain.

Cutaneous perception refers to the perception of an object applied onto the skin in absence of any movement. It includes temperature, pressure, and pain. Kinesthetic perception (perception of movement) is based on the deformation of mechanoreceptors in muscles, tendons and joints, such as in the hand-arm-shoulder system (Gentaz, 2003). It can also rely on the efferent copy of the movement that is being performed. The kinesthetic perception provides feedback on the current position, and movement of body parts. Kinesthetic feedback can also be referred to as proprioceptive feedback.

Haptic perception has been defined as the combination of cutaneous and kinesthetic perceptions in a complex manner across space and time (Gentaz, 2003). It is a dynamic process that combines the cutaneous perception with movement, for instance when actively exploring an object or a pictorial space. Cutaneous perception is referred to as "passive", and haptic perception as "active" (Lederman & Klatzky, 2009). In the field of Human-Computer Interaction, "Haptics" has recently been used to describe any form of interaction involving touch.

#### *Point(s) of contact*

In this chapter, a "point of contact" refers to the location of the map that is currently being explored, either indirectly, by the means of a pointing device, or directly, by the user's hands or fingertips. The number of points of contact that can be simultaneously tracked depends on the technology being used. For example, when exploring a Digital Map with a regular mouse, users rely on a single point of contact. On the contrary, several points of contact are simultaneously available when users directly explore a Physical Map with both hands.

### *Tactile exploration*

Many cognitive processes are demanding during tactile exploration of a map. For instance, vision allows the immediate estimation of the relative position of two objects (distance and orientation between them) on a map. This is not the case during tactile exploration of a map. Here the same estimation is based on three serial processes including the exploration of the whole map, the localization of the items displayed on the maps, and the comparison of their relative position. These processes are based on successive one-handed or simultaneous two-handed movements. For instance, it has been observed that visually impaired people use one finger as a fixed reference point when exploring adjacent parts of a tactile map (Wijntjes, van Lienen, Verstijnen, & Kappers, 2008b). Obviously, when constraining tactile exploration to a single point of contact, spatial integration is even slower and more complicated, and leads to more significant cognitive issues (Heller, 1989; Loomis, Klatzky, & Lederman, 1991).

### *Tactile reference frames*

Obviously a map has its own reference frame, but the user exploring the map with the hands will build own mental reference frames. In psychology, a dissociation has been accepted between "egocentred" and "allocentred" frames of reference (see e.g. Lederman and Klatzky, 2009). In the egocentred reference frame, coordinates of items, and hence distances and directions, are specified relative to the person who is exploring the map. In the allocentred reference frame, coordinates are attached to an external landmark within the display.

Human vision provides both egocentred and allocentred coordinates simultaneously because objects are perceived in relation to the retina but also in relation to each other. This process is automatic and immediate. It is harder to construct both egocentred and allocentred reference frames by touch. Touch is by essence egocentred, and primarily attached to the part of the body that is in contact with the map (usually the hand). Then the origin of the egocentred reference frame is the user itself, and it is more reliable if the user does not move (i.e. change of location relative to the map). In order to build an allocentred mental representation of the map, it is mandatory to explore many items successively. The origin of an allocentred reference frame must be a static tactile landmark on the setup. Finally, because touch is a serial process that relies on direct contact with the various elements of the map, spatial integration by touch, regardless whether it is egocentred or allocentred, needs time.

## I.3   Motivation and method for the classification

Since the first accessible interactive map prototype has been proposed by Parkes (1988), many researchers have designed interactive maps using a variety of technologies and interaction techniques. The current book chapter is an attempt of structuring the work that has been conducted on accessible interactive maps until today. Of course, we acknowledge earlier surveys of interactive maps, which fed the current work. For instance, Zeng and Weber (2011) classified interactive maps for visually impaired people in four groups: 1) "virtual acoustic maps" are entirely based on verbal and non-verbal audio output (for instance, the user interacts by tapping on a tablet which then produces audio feedback); 2) "virtual tactile maps" make use of haptic devices (e.g., a force feedback joystick or mouse); 3) "braille tactile maps" are based on the use of dedicated raised pin displays; and finally, 4) "augmented paper-based tactile maps" use a raised-line map as overlay over a touch-sensitive display combined with audio output. More recently, Kaklanis, Votis, and Tzovaras (2013) presented an overview of accessible interactive prototypes. This survey also contained many accessible map prototypes. In the current survey, we propose a broader classification, including for instance Tangible Maps for the visually impaired, and covering more references to research projects.

Except in (Weber & Zeng, 2011), we did not observe any attempt to more precisely define a terminology to refer to the different types of accessible interactive maps. Various names have been chosen in different publications, and there is rarely an explanation about the choices that have been made. Several authors used the term "audio-tactile maps" (Jacobson, 1998a; Miele, Landau, & Gilden, 2006; Paladugu, Wang, & Li, 2010; Parente & Bishop, 2003; Wang, Li, Hedgpeth, & Haven, 2009). In general, but not systematically, this term refers to maps that are based on touch-enabled devices with audio output. Frequently, but not systematically, "audio-tactile maps" are based on raised-line map overlays augmented with audio output. The term "virtual tactile maps" has also been used to refer to the same kind of prototypes (1999). Some authors referred to different prototypes, based on haptic devices and audio output, as "haptic soundscapes" (Golledge, Rice, & Jacobson, 2005; Lawrence, Martinelli, & Nehmer, 2009). In fact, there is no overall agreement about the nomenclature of accessible interactive maps.

In this book chapter, we present a comprehensive summary of accessible interactive maps for visually impaired people. We cover a much larger design space than in previous work, and propose a classification and terminology. We performed an exhaustive search with the aim of covering as many relevant publications as possible using different scientific databases (ACM Digital Library, SpringerLink, IEEE Explorer, and Google Scholar). Then, we applied different criteria: first, we only considered interactive maps that were designed specifically for visually impaired people. Second, we only included publications in journals or peer-reviewed conferences. Third, publications that proposed concepts without any implementation were discarded. Fourth, if several papers were published on the same prototype, we only considered one publication for each. Exceptions were made if the map prototype has had major changes between successive papers.

## II. Classification of interactive accessible maps for visually impaired users

In the current classification, we distinguish "Digital Interactive Maps" whose display is purely digital (i.e. none of the elements of the maps is embedded into a physical object) from "Hybrid Interactive Maps" whose displays are both digital and physical. In this chapter we did not include maps that are purely physical, i.e. that do not make use of any interactive technology (such as raised-line maps). In the first section we present Digital Interactive Map prototypes, according to the artefact used to explore the map (finger, joystick, etc.) In the second section we present prototypes of Hybrid Interactive Maps and classify them into three subcategories: Interactive Tactile Maps, Tangible Maps, and Refreshable Tactile Maps.

### II.1 Digital Interactive Maps (DIMs)

DIMs can be displayed on a screen or projected onto a surface. They can be explored with one or many points of contacts, which are either direct (e.g. fingertips) or indirect (e.g. the cursor of a mouse or keyboard). DIMs provide auditory feedback and/or tactile feedback (texture, relief, pressure, and/or vibration), according to the finger or cursor position. Most of the regular input devices, such as keyboards or joysticks, do not provide additional tactile feedback. In that case, only audio feedback is provided. More recent input devices can provide additional tactile feedback (e.g. a mouse with braille cell or a force-feedback joystick).

### II.1.1 Regular 2D pointing devices

#### *Keyboard*

The keyboard is a standard device for both sighted and visually impaired people. It has been used in several DIM prototypes (Bahram, 2013; Parente & Bishop, 2003; Simonnet, Jacobson, Vieilledent, & Tisseau, 2009; Weir, Sizemore, Henderson, Chakraborty, & Lazar, 2012; Zhao *et al.*, 2008).

iSonic (Zhao *et al.*, 2008) is a tool for the exploration of georeferenced statistical data (thematical maps). The map is divided into a 3x3 grid. Each cell is mapped to one key of the numerical keypad. By pressing one of the nine keys, users can retrieve data associated with the corresponding region. The arrow keys enabled users to navigate within the map. The keyboard can also be used for zooming. The authors conducted a user study showing that most of the participants used the 3x3 keys to navigate the map, while the arrow keys were used to answer specific questions such as finding the adjacent regions. Some participants managed to understand the overall layout of the map using the 3x3 navigation keys. This map prototype also worked with a touchpad.

Delogu *et al.* (2010) compared two techniques to navigate a sonified map: a keyboard and a tablet. Users were asked to move the cursor across regions, and listen to auditory information about the current location. Then they had to identify the map that they have previously explored among a set of different maps. The results did not show any significant difference in the map recognition task depending on the type of input device used for map exploration (keyboard or tablet). Both input devices enabled users to build an effective cognitive map. However, the results showed that tablet users were more exhaustive than keyboard users, i.e. they explored a higher number of regions. In addition, tablet users changed the direction of exploration more often and were faster. Delogu *et al.* (2010) concluded that the absence of a reliable haptic reference frame and the step by step displacement when using the keyboard demanded a greater cognitive load for integrating the map configuration.

In many other digital maps, keyboard use was limited to additional functions (i.e. command selection) rather than spatial exploration. For instance, it has been used to change the map heading (Simonnet *et al.*, 2009) or to enter commands such as zooming or scrolling (Bahram, 2013).

#### *Joystick*

Picinali *et al.* (2014) implemented a device that used a regular joystick for navigating a virtual environment. The virtual environment represented a corridor leading to a few rooms as well as various objects (doors, windows, elevators) and 3D virtual sounds (music, voices, etc.) Footstep noises were played every 50 cm, and finger snapping noises could be triggered by the user at any time to determine the position of objects by listening to the echoes. The navigation speed depended on the pressure applied to the joystick. Results of their user study showed that participants were able to build correct mental representations of the environment, and that these mental representations were similar to those acquired through actual navigation in the real environment. Similarly to the keyboard, regular joysticks do not provide any feedback other than visual regarding the position of the cursor. In addition, the cursor movement is relative to the last cursor position, which is hard to track without vision. Although the results by Picinali *et al.* (2014) were encouraging, joysticks have rarely been used in accessible digital maps.

### *Tangible pointing devices*

In this subcategory we refer to the use of objects that have been used to move a cursor over the map. We do not refer to maps whose representations are embedded into several tangible objects (see 0). In the specific context of the current chapter, we consider the computer mouse as a tangible pointing device.

Only few projects were based on using a regular mouse. Earth+[1] is a project developed by the NASA where the user moves the mouse within the map. The visual image was transcribed into auditory feedback, with different colors corresponding to different notes. These notes were then played corresponding to the current cursor position. To our knowledge the device has not been further developed since 2009, and remained in beta version. In addition it has not been evaluated with visually impaired users. In fact, it appears that mouse is rarely used by visually impaired people because the feedback concerning the cursor position is primarily visual. In addition, when using a mouse, distortions appear between perceived and real distances (Jetter, Leifert, Gerken, Schubert, & Reiterer, 2012). Finally, the mouse can be lifted up and moved elsewhere while keeping the pointer position stationary (Lawrence *et al.*, 2009), which generates disorientation when using it in absence of vision (Golledge *et al.*, 2005; Pietrzak, Crossan, Brewster, Martin, & Pecci, 2009).

Milne *et al.* (2011) used a pen-based digitizer tablet with a stylus that enabled users to control the position of the cursor within the map. Orientation was controlled by rotating the stylus (a tactile cue on the stylus indicated the forward direction). Daunys and Lauruska (2009) used a similar approach. The users explored the map by moving a pen over a digitizer tablet, and non-speech sounds were played accordingly. Brittell, Young and Lobben ( 2013) also used a similar apparatus to present a choropleth map to visually impaired users, including basic spatial analysis tools. Depending on the position of the stylus, different queries were sent to a spatial database to provide the user with up-to-date content. Different sounds were played to inform the users about the population density as well as to indicate when they were near a border or outside the map. Pielot *et al.* (2007) used a toy as a "virtual listener" that was moved by the user over a map. A camera placed above the map was tracking the object's position and orientation, and rendered auditory output according to its position.

### II.1.2 Pointing devices with additional somatosensory feedback

Visually impaired people are used to both auditory and tactile cues. The afore-mentioned prototypes provide only auditory feedback, which may limit their ability to convey information. Many DIM prototypes have used pointing devices with additional force or cutaneous feedback.

### *Pointing devices with force feedback*

Force feedback devices mechanically produce a force that is perceived as a kinesthetic feedback by the user (El Saddik, Orozco, Eid, & Cha, 2011). Various force feedback devices have been used for the exploration of DIMs by visually impaired people including computer mice, devices with handles, gamepads and joysticks.

---

[1] http://prime.jsc.nasa.gov/earthplus/ [last accessed September 29th 2016]

As mentioned before, regular computer mice are difficult to use in absence of vision. Force-feedback mice provide additional tactile feedback that may be helpful (Campin, McCurdy, Brunet, & Siekierska, 2003; Lawrence *et al.*, 2009; Parente & Bishop, 2003; Rice *et al.*, 2005; Tornil & Baptiste-Jessel, 2004). For instance, in the map prototype by Rice *et al.* (2005) a haptic grid overlay, and a haptic frame were rendered by a force feedback mouse. Moving the mouse over the grid produced force-feedback, and allowed users to keep a sense of distance, scale, and direction. The haptic frame around the map served as a barrier to present the map outline. Rice *et al.* (2005) reported that the frame was very helpful for the users as it works as a reference frame (see section I.2.2). However, Lawrence *et al.* (2009) observed that users encountered problems regarding spatial orientation with such a device even if a grid was provided.

Other prototypes used gamepads (Parente & Bishop, 2003; Schmitz & Ertl, 2010) or joysticks (Lahav & Mioduser, 2008; Parente & Bishop, 2003) with force feedback. Both are affordable and available as mainstream products. They generally possess a small number of degrees of freedom and moderate output forces (El Saddik *et al.*, 2011). In the BATS prototype (Parente & Bishop, 2003), the input device provided slight or large bumps when the cursor moved across boundaries, as well as vibrations when the cursor moved over a city. Schmitz and Ertl (2010) used the vibrations of the gamepad to indicate when the cursor was in proximity of a street. Users could navigate the map by moving the analog sticks of the gamepad. In the prototype of Lahav *et al.* (2008), users could navigate a virtual environment using a force-feedback joystick. The environment used in their study represented a room with several elements such as doors, windows, and pieces of furniture. Footsteps noises were played during the displacement as well as other sounds (tapping, bumping, names, etc.). Results showed that participants were able to understand the spatial configuration of the room, but also to rely on the build mental map to successfully perform Orientation and Mobility tasks in the real environment.

Other haptic devices rely on a handle that can be moved and eventually rotated in space and that allows interaction in three dimensions. Both the Geomagic Touch X$^2$© (formerly the Sensable Phantom Desktop) and the Novint Falcon$^3$© are tensioned cable systems, i.e. the handle is moved in different directions by several actuated cables (El Saddik *et al.*, 2011). The user grasping the handle can sense the force that is produced by the device. The Geomagic Touch X provides six degrees of freedom (DoF), i.e. the possibility to vary position and orientation along the three spatial axes, as well as 3D force feedback. The Novint Falcon allows only three DoF (El Saddik *et al.*, 2011). Several DIMs have been implemented with these devices. The prototype by De Felice *et al.* (2007) allowed the exploration of indoor environments as well as complex geographical areas. Each element of the map (doors or rivers for example) was associated with a specific haptic feedback. Users could select the content that they wanted to display as well as change the scale and level of details of the map. HaptiRiaMaps (Kaklanis, Votis, Moschonas, & Tzovaras, 2011) was an open-source web application based on OpenStreetMap$^4$ (OSM). It allowed visually impaired users to search for a specific address, and explore the haptic and audio map around the selected address. VAVETaM (Verbally-Assisting Virtual-Environment Tactile Maps; see Lohmann, Kerzel, & Habel, 2010) provided visually impaired users with the name of the elements that they were currently exploring but with an additional description of the local spatial layout (for example, "this is the intersection between road A and road

---

[2] http://geomagic.com/en/products/phantom-desktop/overview [last accessed September 29$^{th}$ 2016]
[3] http://www.novint.com/index.php/novintfalcon [last accessed September 29$^{th}$ 2016]
[4] https://www.openstreetmap.org [last accessed September 29$^{th}$ 2016]

B"). SeaTouch (Simonnet *et al.*, 2009) allowed blind sailors to prepare a journey in a DIM providing haptic feedback, text-to-speech output (TTS), as well as ambient sound.

Iglesias *et al.* (2004) worked with the "GRAB" interface. This device creates 3D force-feedback too, but in contrast to the previously mentioned devices it has two distinct handles. Two fingers, either of the same or different hands, are placed in two thimbles onto which two independent force-feedbacks are applied. Several applications were developed, one of which being a city-map explorer application. Observations confirmed that using a second finger "can be vital as an "anchor" or reference point" (see Points of contact in I.2.2)

### *Pointing devices with cutaneous feedback*

Input devices can also be augmented with cutaneous feedback. The more common devices include mice with an array of pins, in which the pins move up and down according to the cursor location. For instance, the VTPlayer by VirTouch was such a tactile mouse with two 4x4 arrays of pins. These arrays were located under the index and middle fingers, and were actuated according to the cursor location. We identified one project (Jansson, Juhasz, & Cammilton, 2006) that made use of a mouse with an array of pins that were actuated according to outlines and textures. Jansson *et al.* evaluated this device with 60 sighted blindfolded participants comparing different map representations. They recommended that shapes should be represented without any filling when using such a device. More recently, Tixier *et al.* (2013) designed Tactos, a system made of two devices: one pointing device moving over the map, and two fixed braille cells placed on another device on the side. Moving the pointing device determined which information was displayed on the braille cells. Different pointing devices could be used, such as a stylus on a digitizer tablet. The Tactos was not evaluated.

Latero-tactile displays have been designed quite recently and provide a different cutaneous feedback. A latero-tactile display deforms the skin of the fingertip with an array of laterally moving pins actuated by miniature motors (Levesque, Petit, Dufresne, & Hayward, 2012; Petit, Dufresne, Levesque, Hayward, & Trudeau, 2008). It can evoke perceptions like dots, grating (waves) and vibrations. In a first study (Petit *et al.*, 2008), the latero-tactile device was used to explore a map with two levels of information: one for the continents and the other for the location of five civilizations. Users could switch between these two levels by pressing a key. In this study, the device was mounted on the Pantograph (Campion, 2005), a haptic device that allows 2D-movement over a limited surface. In another study, Levesque *et al.* (2012) observed nine blind users exploring a concert hall seat plan with a latero-tactile display mounted on a movable carrier. The carrier measured the absolute location of the device. These studies showed that visually impaired people as well as blindfolded sighted people could successfully explore the maps with the device. However, because this device is still a lab prototype, only few projects made use of it.

### II.1.3 Finger-based exploration

Finger-based exploration, which we also call "direct exploration", has been used in many research projects. In that case the input device is one of the user's fingers, onto which the position of the cursor is directly mapped. Feedback can be auditory or tactile (vibrations). The location of the user's fingers can be tracked using a touch enabled device (e.g., digitizer tablets, smartphones, tablet or tabletops) or a camera.

### Finger tracking on touch-enabled devices

Because smartphones, tablets and digitizer tablets are mainstream consumer devices, many accessible map projects have used this kind of device. Smartphones and tablets moreover present the advantage of being usable in mobile situations. Large tabletops are more expensive and have been used less frequently. Some projects did not specify the type or size of touch device and might therefore function with different hardware (Bahram, 2013; Parkes, 1988).

Several projects relied on touchpads (i.e. single input devices that are commonly used by graphic designers and that enable interaction using a finger or a pen). The Jacobson's prototype (1998) allowed visually impaired people to explore auditory maps by pressing specific areas on a touchpad (north, west, south, east, and zoom buttons). Verbal descriptions, ambiant sounds (such as traffic noise), and auditory icons were played during exploration. Five visually impaired and five blind people evaluated this device. All of them were able to use it and found it simple, satisfying and fun. The iSonic project (Zhao *et al.*, 2008) also relied on a touchpad but with an additional keyboard for exploring chloropleth maps. When tapping on a region, one of five violin pitches was played according to the chloropleth's value within that region. Using iSonic, seven participants were able to solve complex tasks. Participants found the prototype easy to use and helpful.

Recently, smartphones have been used. TimbreMap (Su, Rosenzweig, Goel, de Lara, & Truong, 2010) was a sonification interface for exploring floor-plans of buildings. Stereo auditory feedback indicated the corrective action needed to follow a path on the screen, or the direction that the finger was drifting towards. Six blind participants managed to identify simple shapes with 80% accuracy. One participant performed a supplementary task in which he was able to find all points of interest on two maps of different complexities. TouchOverMap (Poppinga, Magnusson, Pielot, & Rassmus-Gröhn, 2011) provided a basic overview of a map layout, by giving vibrational and vocal feedback when the finger passed over a map element (e.g. streets or buildings). The evaluation, performed with eight blindfolded sighted users, showed that all participants acquired a basic understanding of the "zoomed-out" and "zoomed-in" version of a map, even though they found the "zoomed-out" condition difficult.

Several projects relied on the use of tablets that are larger than a smartphone but smaller than a computer screen. Simonnet *et al.* (2012) designed a tablet application with auditory and vibrational feedback. Carroll, Chakraborty, and Lazar (2013) described the design of a weather map of the USA for visually impaired users. When the user moved his/her finger over the tablet, a TTS engine pronounced the name of the states and different musical notes were played to convey temperature values. When the user was touching a state containing a minimal or maximal temperature, the device vibrated. Double-tap was used for zooming. The same design was presented by Lazar *et al.* (2013). Another project called Open Touch/Sound Maps (Kaklanis, Votis, & Tzovaras, 2013a) provided visually impaired users with access to OpenStreetMap data using TTS synthesis, vibration feedback and sonification. Vibration and TTS description were executed when the user's finger touched a point of interest. Besides, 3D auditory cues were provided to indicate the distance between the current finger location and the next crossroad.

In the framework of the BATS project, Parente and Bishop (2003) developed two prototypes to enable a blind student to explore a map of Great Britain. The user could explore the map either by moving the on-screen cursor with a mouse or a trackball or by using a touch-screen directly. TTS and spatialized auditory icons were provided. An informal evaluation with one user revealed that he preferred using a trackball to the touchpad. However, an informal evaluation with one user only was not suficient to draw conclusions.

Few projects have been developed on tabletops. Kane, Morris, *et al.* (2011) designed three interaction techniques for maps displayed on large touch-screens. In a first bimanual technique called "Edge Projection", locations of map elements were projected onto the x- and y-axes on the left and lower edges of the screen. Users could thus browse the axes to find the names of on-screen targets. When they identified a target, they could drag both fingers from the edges to the interior of the screen in order to locate it. The second technique, called "Neighborhood Browsing", expanded the size of map elements by using the empty area around them. Touching the neighborhood of a map element launched audio feedback pronouncing the element's name. The third technique, called "Touch-and-Speak", combined touching the screen with speech recognition, so that a user could, for example, ask for the list of on-screen targets. Both "Neighborhood browsing" and "Touch-and-Speak" allowed to find a specific target based on spoken instructions. Fourteen blind users compared these three interaction techniques to a regular implementation of Apple's VoiceOver. The results revealed that "Touch-and-Speak" was the fastest technique, followed by "Edge Projection". Furthermore, there were significantly more incorrect answers with VoiceOver than with the other techniques. Users also ranked "Edge Projection" and "Touch-and-Speak" significantly better than VoiceOver. Another tabletop based prototype was designed by Yairi *et al.* to enable users to follow a line on a map (2008). A line was divided into eight equidistant segments. When the finger followed the line, the successive notes of an octave ('do re mi fa sol la si do') were played according to the segment being touched. The authors asked four blind people to explore a map with streets, and then find a route without assistance. Using this technique, users were able to know the length of each route segment being touched. Interestingly, they were able to anticipate the next crossing when a route was made of many lines. Finally, all participants reached the goal of finding a route, even though they sometimes made wrong turns or felt lost.

In the Tikisi project (Bahram, 2013) the touchscreen type and size was not specified. Users could touch a map and hear the name of the element under their finger (country, road, etc.). They could also zoom in and out, scroll, or select a location. Additional input techniques (keystrokes and speech recognition) were used, and TTS was provided as output. Twelve visually impaired people used the application and reported surprise, enjoyment and interest. However the study did not report any result about success.

### *Camera-based finger tracking*

Cameras (including webcams, embedded smartphone cameras, stereo cameras or motion-capture systems) can be used to detect one or many user's finger(s). In contrast to touch-enabled devices, it is possible to identify each finger. On the other hand, these systems are subject to technical challenges such as lighting conditions and occlusion.

The "KnowWhere" system (Krueger & Gilden, 1997) was composed of a camera mounted above an illuminated table. A sound was played each time the user's fingertip passed over an element. Users could also enlarge a specific area in order to access a more detailed map. Another prototype developed by Schneider and Strothotte (1999) allowed the exploration of an urban area as well as learning a route. The feedback was based on both speech and sound. Seisenbacher *et al.* (2005) used a similar set-up, but with a color marker attached to one finger and tracked by a camera mounted above a tabletop. When the finger entered an interactive zone, the corresponding sound was played. AccessLens (Kane, Frey, & Wobbrock, 2013) enabled visually impaired users to interact with paper documents including elements with labels. The camera placed above the table was used to decode the labels and recognize the user's gestures. Users were able to retrieve the names of the elements, but also to use a gesture or voice command menu. During the evaluation, five blind users were asked to explore a diagram, a map of the USA, or a table presenting poll results. Qualitative results showed that the participants were highly satisfied with the system and the interaction modes, even if they faced some interaction issues (when placing both hands on the document for example).

Bardot *et al.* (2014) used a motion tracking system to track a smartwatch during map exploration. Motion tracking allowed both 2D (exploration) and mid-air gestures (filtering). The smartwatch provided sound and vibrational feedback, in addition to different filtering commands. A two-step guidance function based on vibrations helped users to find specific targets. In a follow-up study, Bardot *et al.* (2016) used a similar set-up to track the user's dominant finger. The maps were composed of several areas, each area being associated with a name and quantitative data. Three exploration techniques were designed: the "Plain" exploration technique simply provided auditory feedback when the finger was entering an area; the "Filter" technique relied on filter selection on the smartwatch and enabled the user to select which data to display; the "Grid-Filter" technique combined the "Filter" with the use of a virtual 3x3 grid that the user could explore using mid-air gestures. The evaluation, including 12 visually impaired participants, compared the exploration of a regular raised-line map to the exploration of a digital map with these three different techniques. It showed that the exploration of a digital map, without any tactile cues, is possible. It also showed that the Grid-Filter technique is efficient for data selection or comparison tasks.

### II.1.4 Summary and conclusions

In this category we discussed different prototypes that provide visually impaired people with access to digital maps, i.e. maps whose representation do not rely on any physical object as elements of the map. The term "virtual", which is sometimes used, is appropriate, but should not be used in opposition to visual maps. In comparison to "visual" maps, these maps could specifically be called "auditory" or "somatosensory" maps. We chose to use the term "digital" in order to incorporate in the category all the prototypes with auditory and/or somatosensory feedback (in many prototypes both feedback are used simultaneously), but also to stress the opposition with hybrid maps that combine a physical display (e.g. raised lines maps, small scale models, or refreshable displays) with a digital one.

As we showed, only few accessible interactive map prototypes made use of a keyboard. A keyboard is very well adapted to text entry, but is less adapted to 2D pointing and hence to exploration of spatial data. In the context of accessible maps, keyboards have mainly been used as a complementary text and command input device. For instance, it has been used to change the map heading (Simonnet *et al.*, 2009) or to enter commands such as zooming or scrolling (Bahram, 2013).

Joysticks are based on the relative displacement of a cursor, which works well with visual feedback. In absence of vision, the location of the cursor must be perceived through specific auditory or somatosensory feedback. The localization of the cursor then engenders an additional cognitive task that must be performed by the user. In the prototype of Picinali *et al.* (2014), participants managed to explore a virtual environment with a joystick. But the environment was rather simple (a corridor and a few objects), and it is uncertain if a joystick could be used to explore more complex environments.

In contrast to keyboard and joystick, a tangible pointing device allows the user to point at specific locations on a digital map. The location of a tangible pointing device is perceivable by touch and is trackable with a camera or a touch-screen device. Then no dedicated hardware is required, and if the object does not provide additional feedback (e.g. vibrations) it can be very cheap. Indeed, affordable mainstream devices such as mice, pens and toys can be used with a cheap camera (Pielot *et al.*, 2007).

Many tangible pointing devices provide additional cutaneous or haptic feedback. It has been shown that such feedback may increase usability. For instance, the augmentation of a computer mouse with force feedback that indicates map boundaries and edges of a grid helped users to maintain spatial orientation (Rice *et al.*, 2005). Other pointing devices with tactile displays (e.g. braille cells or latero-tactile display) also enhance map exploration, for example by enabling users to switch between different levels of information (Levesque et al., 2012). However, these devices are lab prototypes, and hence are not available for visually impaired users. Specific haptic devices, such as the Geomagic Touch X, are on the market, and also provide sensory cues that enhance accessibility. For instance, they have been proved efficient for exploring sea maps by blind users (Simonnet, Vieilledent, Jacobson, & Tisseau, 2011). However, they are more expensive than regular pointing devices.

Touch-screens, tablets and smartphones have been frequently used for non-visual map exploration because they are widespread and affordable. They provide a physical surface onto which the map is displayed, which helps the user to select a reference frame (e.g. one corner of the touchscreen), and quickly perceive the spatial extent of the map (it usually corresponds to the size of the touchscreen). In addition, tablets and smartphones can be used in mobile situations, which may be very important for visually impaired users. However, touchscreens do not provide cutaneous feedback associated to the elements that are displayed on the map. They are *per se* input devices with visual display (Buxton, 2007). In the absence of tactile cues, it is difficult to understand the spatial layout and estimate relationships between elements of the map. It is then necessary to provide additional non-visual feedback for visually impaired users.

Camera-based digital maps provide the same advantages and drawbacks as maps based on touch-enabled devices. However, they highly depend on lighting conditions and occlusions, which make them less usable by visually impaired users, especially when they are used in mobility. Finally, maps based on motion capture devices offer enhanced opportunities because they are adapted to the capture of 3D gestures. They can provide mid-air (Bardot et al., 2014, 2016) or complex gesture recognition, but they are much more expensive and need accurate calibration as well as a fixed setting. It is interesting to note that even though some force-feedback devices (e.g. Geomagic Touch X) can detect 3D gestures, no prototype took advantage of this capability.

In conclusion, digital maps can be explored using a variety of input devices, as well as with touch and gesture interaction. Digital maps present many advantages. Depending of the input device that is used, a digital map prototype can be very cheap (e.g., Su *et al.*, 2010) and used in various contexts including home, the workplace, or school. In addition, digital maps can be dynamically updated, and therefore support operations such as panning and zooming. However, the absence of tactile cues during non-visual exploration is an important drawback that must be compensated.

## II.2 Hybrid Interactive Maps (HIMs)

As we previously mentioned Hybrid Interactive Maps (HIMs) are made of a digital and a physical display. There have been many prototypes in the literature relying on different physical displays. We classified HIMs into three categories according to the type of physical display that was used.

### II.2.1 Interactive Tactile Maps (ITMs)

With the term "Interactive Tactile Maps" we refer to physical displays that are tactile maps. In contrast with tangible or refreshable maps, which we will introduce hereafter, tactile maps are static. The content cannot be dynamically updated. The only way to change the map view (zoom or pan for instance) is to change, or erase and redraw the physical display. In the category of tactile maps, raised-line maps have been extensively used, but more recently 3D printed maps have emerged. Two different technologies have then been used to track the user's finger(s) over the tactile map: touch-enabled devices or cameras.

#### *Interactive Raised-Line Maps*

Raised-line maps have proved beneficial for spatial learning by visually impaired people for many years, and are thus a familiar tool for most of them (Ungar, 2000). Several prototypes relied on this observation, and aimed at making raised-line maps interactive. In those maps, a raised-line overlay is placed over a touchscreen device that allows the detection of touch inputs through the overlay. Users perform taps or double-taps on any interactive element of the overlay, which produces speech, sound or vibrational feedbacks.

Parkes (Parkes, 1988) was the first to design an Interactive Raised-Line map with an overlay placed over a touchscreen. Even though the technical aspects of his NOMAD prototype were not precisely described, Parkes envisioned that gestural interactions (for example "touching two points for direction and distance interaction") combined with appropriate auditory feedback could greatly enhance the use of tactile maps. Since then, several research projects have been developed along this direction. The prototype described by Weir *et al.* (2012) presented weather forecast of the USA. When the user selected a state by tapping on it, a sound was played whose pitch encoded temperature (the higher the pitch, the higher the temperature). The prototype of Senette *et al.* (2013) provided the names of the streets using a TTS engine. It was enriched with non-speech sounds and vibrations that were activated when the user touched pedestrian zones. Hamid and Edwards (2013) presented an interactive raised-line map placed on a turntable. When rotating the turntable, users could explore a route from an egocentered point of view (i.e. from the traveler's perspective). In addition to the satisfaction provided by the map rotation, users reported that sounds attributed to ground textures were especially useful.

As described above, Parkes envisioned using gestural interaction in interactive raised-line maps. To our knowledge, only Brock *et al.* (2014) implemented gestural interaction in such a map type in order to access distances between two points or to get additional information about map elements.

Mappie (Brule *et al.*, 2016) was an extension of this previous prototype. Mappie's overlay used different colors in order to be accessible not only for blind, but also for low vision and sighted people, which enabled them to collaborate. In addition, a menu bar was included to enable users to choose between different types of spatial content. Then, Brulé *et al.* (2016) developed the MapSense prototype that relied on the same interactive raised-line maps, but provided fourteen additional conductive tangibles. Some tangibles could be filled with scents, such as olive oil, smashed raisins or honey, thus creating a real multi-sensory experience including olfactory cues.

Because they are usable, Interactive Raised-Line Maps are being used in the wild. ABAplans[5] is a project initiated by the Engineering School of Geneva (Ecole d'Ingénieurs de Genève). The device is based on a raised-line map overlay placed over a mono-touch sensitive surface. It includes a specific map editor, and allows users exploring a city map, finding a specific place, preparing journeys, and learning about public transportation. It is currently distributed by the AudioTactile association. The company ViewPlus commercializes a similar device called IVEO[6]. It includes a mono-touch sensitive surface and an editor for sketching raised-line maps and drawings. It is possible to purchase additional software that allows creating raised-line images from PDF or scanned documents using optical character recognition. In both cases, users must own the mandatory equipment to print raised-line maps.

Weir *et al.* (2012) compared the usability of an Interactive Raised-Line Map to the usability of a Digital Interactive Map based on a touchscreen or a keyboard. The evaluation showed that the participants preferred the Interactive Raised-Line Map. A few studies also compared the usability of Interactive Raised-Line Maps to regular (non-interactive) raised-line maps (see e.g. Wang et al., 2009). Six blind users raised positive comments on the clarity of information that was provided. They found that the device was easy to use and helpful for pedestrian navigation. In a follow up study, Wang, Li, and Li (2012) compared an Interactive Raised-Line Map with a tactile map without any textual information. They observed that users preferred the interactive map. Furthermore, they observed that the interactive map was quicker in 64% of all cases for identifying start and end points of a route, but not for route exploration. However this comparison should be considered with precaution because, users spent most of the time listening to the audio output in the interactive condition, whereas the tactile map did not contain any textual information. Brock *et al.* (2015) compared the usability of a regular raised-line map to an Interactive Raised-Line Map that displayed the exact same content. The evaluation included twenty-four blind participants that were required to explore an unknown neighborhood. The study assessed the time needed for exploration, the accuracy of the spatial learning, and the satisfaction of the users. The results showed that interactivity significantly shortened exploration time and increased user satisfaction.

While the previous studies have been done in a laboratory setting, other studies have been performed in a classroom scenario. Brulé *et al.* (2016) conducted a formative study with the Mappie prototype over several months in a special education center. The results were promising and provided insights in the design of accessible interactive maps, such as using other sensory modalities (i.e. olfaction and taste) to foster learning. They also recommend using additional interactive objects to support storytelling. The MapSense prototype has also been used during two classes made by a locomotion trainer and a specialized teacher. The authors observed that the device triggered strong positive emotions and stimulated learning as well as creativity of the visually impaired students.

---

[5] http://www.audiotactile.ch/ [last accessed June 3rd 2016]
[6] https://viewplus.com/product/iveo-hands-on-learning-system/ [last accessed June 3rd 2016]

### *ITMs with a camera-based finger tracking*

Interactive Raised-Line Maps usually combine a touch-enabled device with a raised-line overlay. This is functional because touch inputs can be detected through the overlay. Touch-enabled devices cannot be used with thicker representations such as 3D prints. Then another way of tracking the user's fingers is to use a camera. With the Tactile Graphic Helper (Fusco & Morash, 2015), a visually impaired user can place a tactile map on a regular table and then interact with it. The camera, placed above the tactile drawing, recognizes the layout and tracks the user's fingers. The user can point at elements and ask for information. The Tactile Graphic Helper aimed at allowing visually impaired students to discriminate tactile symbols (texture, Braille labels, etc.) without requiring the help of a sighted person.

Sullivan and Picinali (2014) used another technology, the Leap Motion©, to detect pointing movements towards elements of a tactile map placed over an inclined table. The user had to perform a distinct pointing gesture to select a specific element. Two types of auditory feedback were given, either to provide users with detailed information about specific elements, or to guide them along a route. Spatial sounds were also played, as, for example, sounds of flowing water when a river was selected.

Götzelmann and Pavkovic (2014) designed interactive 3D-printed maps. The production of 3D maps was automatic so that visually impaired users could make them without assistance. Once printed, the fingers were tracked with a smartphone held above the 3D map. The application identified the map with a printed barcode, and, in addition, helped the user to correctly hold the smartphone over the map. With the other hand, the user was free to explore the map, and received auditory feedback when pointing at elements.

### *Erasable Tactile Map*

Recently, an interesting prototype based on 3D printing has been designed by Swaminathan *et al.* (2016). LineSpace is a platform that includes a movable 3D printer head mounted over a drafting table. The system could print and erase spatial content based on vocal commands and deictic gestures. The gestures were detected by a camera tracking the markers affixed to one users' finger. It provided visually impaired users with the possibility to dynamically draw and explore spatial content. Among the different applications described in the paper, one allowed users to search for real estates within a city map. Once the map was printed, the user could retrieve additional information about a particular estate. For rescaling the map or exploring a new part, the system printed a new map on a blank part of the drawing table, which enabled the user to switch back and forth between two different views. The system can also remove elements that have been previously printed by using a "scraper".

## II.2.2 Tangible Maps

Tangible user interfaces combine physical objects with digital data, and thus enable interaction with the digital world through the use of physical artefacts (B. Ullmer & Ishii, 2000). As already mentioned, we make a distinction between digital maps that are based on a tangible pointing device (see II.1.1) and tangible maps *per se*. Tangible maps are made of several physical objects that represent map elements and allow bimanual exploration. Users can also manipulate the tangible objects in order to (re)construct or edit the map. On the contrary, tangible pointing devices (such as computer mice, pens or toys) do not represent any element of the map but only serve as a pointing device.

Schneider and Strothotte (2000) designed a prototype that enabled visually impaired students to independently construct an itinerary using building blocks of various lengths. The system indicated the length and orientation of the next building block that had to be placed. The user's dominant finger was tracked during exploration of the virtual map, and guided along the route. More recently, Ducasse *et al.* (2016) designed a novel type of physical icons, called Tangible Reels, which were used to render map points (cities, bus stops, etc.) and lines (streets, rivers, boundaries, etc.) tangible. The Tangible Reels were made of stable objects combined with a retractable reel. The user was guided by audio instructions to correctly place and link the objects on a tabletop, until the whole map was constructed. The user could then retrieve the name of the elements with a pointing gesture above the objects and the links. The evaluation, conducted with eight visually impaired users, showed that the interface enabled the construction and exploration of maps of various complexities (up to twelve objects), in a short amount of time (24 seconds on average to place and link an object), and with very few errors (283 out of 288 objects were correctly placed during the whole evaluation).

### II.2.3 Refreshable Tactile Maps

Refreshable tactile maps refer to maps that are physically rendered (using a matrix of pins for example), and that the system can dynamically update. Obviously, the devices that intuitively may provide blind users with access to any content, at any time, are refreshable tactile displays. They can represent a complete drawing with many different heights for the relief (also called 2.5D[7]). The display can be refreshed dynamically, and hence allows any update, as well as zooming and panning. Up to now, only one technology has been used: raised-pin displays.

Raised-pin displays rely on pins that are raised mechanically either by electromagnetic technologies, piezoelectric crystals, shape memory alloys, pneumatic systems or heat pump systems (El Saddik *et al.*, 2011). Vidal-Verdú and Hafez (2007) referred to this type of devices as static refreshable displays: they are equivalent to a screen "where pixels are replaced by taxels, i.e. touch stimulation units". Visually impaired people are accustomed to this type of display because they frequently use dynamic braille displays that are based on a similar principle (Brewster & Brown, 2004). Braille displays are used to display textual information. They are made of one or two lines of 40 to 80 cells, each with 6 or 8 movable pins that represent the dots of a braille cell.

Four map prototypes have been designed using a large display composed of actuated pins. These displays present information in relief (2.1D). The user moves the hand across the display to explore the content. Obviously, it is possible to dynamically change the content or to highlight elements by raising or recessing specific pins. Zeng and Weber (2010) used the BrailleDis 9000 tablet which was composed of 7200 pins arranged in a 60x120 matrix, and actuated by piezo-electric cells. Touch sensors allowed the user to provide input to the system (tap or double tap for instance). Zeng and Weber designed a tactile symbol set for displaying different types of information such as bus stops or buildings. In a second publication, they improved the tactile symbol set as well as the prototype, and introduced the ATMap prototype that allowed the users to pan and zoom, but also to share annotations (Limin Zeng & Weber, 2012a). In their study, Schmitz and Ertl (2012) used the HyperBraille display[8], a commercial version of the BrailleDis 9000, to display two types of maps: detailed maps representing buildings or small outdoor areas, and overview maps retrieved from OpenStreetMap. The prototype also provided panning and zooming functionalities. When the user typed in an address or the name of a point of interest, the map was refreshed. Ivanchev *et al.* (2014) also used the BrailleDis 9000 to present routes to visually impaired users. They compared several patterns to help the user visualize the routes by varying the thickness of the lines or making the pins blink for example. An observation made on one subject indicated that the blinking mode was efficient. Panning, zooming, layering and searching functionalities were also implemented. Shimada *et al.* (2010) constructed a display with 3072 raised pins. In addition, this system included a scroll bar that indicated which part of the image was inside the displayed area. This device also contained touch sensors in order to track user inputs.

---

[7] 2.5D has been defined in opposition to 2D (flat and without relief) and 3D (object). 2.1D is sometimes used to refer to relief with just one height. Braille cells are 2.1D devices because the raised pins can reach one position only.

### II.2.4 Summary and conclusions

In this large category of physical maps, we presented map prototypes that include an interactive physical representation of the map. The most studied prototypes are undoubtedly the Interactive Tactile Maps. They have been evaluated in many studies, and altogether, these studies show that maps combining touch devices with raised-line overlay and audio output are definitely usable for acquiring spatial knowledge in absence of vision. In addition, they are now on the market and are being used in many situations like at home or at school. The downside of these devices lies in the necessity of tactile overlays, whose production is time-consuming and relies on tactile document makers. In commercial offers, they are sold as packages for a region or a country. Tactile overlays also limit the usage of the device in mobility, because of physical bulkiness and weight (Klatzky, Giudice, Bennett, & Loomis, 2014). Furthermore, zooming or panning operations are not possible without placing another tactile overlay, which may cause cogntive issues.

It is therefore interesting that some projects aimed at automatically producing tactile overlays either by retrieving spatial data from a Geographic Information Systems, or based on image recognition and segmentation. Campin *et al.* (2003) described a prototype that allowed the user to select a SVG map on a dedicated web site, and then to print it at home with a thermal enhancer. In the "Talking TMAP" project (Miele et al., 2006), the maps were automatically generated and could be ordered online. Different levels of information could be accessed through repeated tapping. Auditory feedback included name, length and spelling of the streets, as well as address ranges. A tactile user interface was provided to facilitate the navigation within the main menu (finding a location, calculating distances, modifying the settings). Wang *et al.* (2009) implemented a system that automatically created interactive tactile maps from map images.

In the same subcategory, we placed 3D printed small scale models that are interactive thanks to a camera that tracks the fingers. 3D-printers are affordable, and can be used to print small neighborhoods, but also small scale models of buildings, objects, etc. They already spread over specialized centers, but to our knowledge, no interactive device has already been used on the wild. Talking small scale models are now used in many public places. They are now designed by a few companies (see e.g. Talking Tactile Campus, Touch Graphics, Inc., USA). They rely on touch sensitive zones (not on camera-based finger tracking) and proved useful for wayfinding. However, they are quite expensive and cannot be updated.

---

[8] http://www.hyperbraille.de/?lang=en [last accessed August 21st 2013]

At the intersection between Interactive Tactile Maps and Refreshable Tactile Maps, LineSpace (Swaminathan *et al.*, 2016) is a lab prototype that can be used to dynamically print and erase interactive maps on a drawing board. But the time required to print a map was not indicated by the authors. It is therefore unclear whether the map can be readily updated or not. Obviously, LineSpace could be used in specific places such as specialized centers but not in mobility.

The second subcategory of HIMs called "Tangible Maps" is promising, but there is surprisingly little work on that topic. This may be explained by the fact that Tangible Maps, although they can be relatively cheap (a webcam, a PC, and a transparent table are enough), still require a large tabletop and several objects, which is cumbersome. Commercial interactive tabletops that identify and track tangible objects also exist (e.g., the Reactable[9]) but are expensive. One of the main advantages of Tangible Maps is that they support the autonomous construction of map, which may help the users understanding and memorizing the spatial layout of the map (Ducasse *et al.*, 2016). Besides, because the map representation can be easily manipulated by the user, it is possible to "refresh" the map by moving the objects or adding and removing objects. It is therefore technically possible to rescale or reposition the map by moving the objects. In conclusion, tangible interfaces are not yet mainstream devices, and cannot be used in mobility. However, we guess that in the near future they should become more frequent in public spaces such as museums, schools, etc.

The last subcategory was called Refreshable Tactile Maps. It includes maps whose physical representation can be automatically updated, which is an outstanding advantage. However, the more advanced devices, namely Raised-pin displays, are still very expensive. In 2007, Vidal-Verdu and Hafez (2007) estimated that a display large enough to be explored with two hands would require more than 75 000 pins, and would approximatively cost 270 000€. In 2012, the HyperBraille display with 60 x 120 pins was sold 50 000€. Hence, current raised-pin displays are generally small, with a low resolution (Limin Zeng & Weber, 2010). However, they can be used in mobility (Limin Zeng & Weber, 2012b), which is very important for visually impaired people.

## III. Discussion

### III.1 Comparison of interactive map devices

As shown above, there is a large variety of devices allowing visually impaired users to explore maps. Each type of devices that we described in the previous sections presents advantages, drawbacks, and limitations. To our knowledge, the different types of devices that we presented in this chapter have not been systematically compared. We believe that doing so could help designers and researchers better identify the advantages and drawbacks of each device depending on the tasks that they intend to support, the context in which the prototype will be used, the sensory capabilities of the users, etc.

At the end of the two previous sections (Summary and conclusions), we already discussed the advantages and limitations of the devices used in terms of cost, availability, and technological limitations. We also mentioned whether the prototypes could be used in mobility, and if they allow readily updating the map display. In the following sections, we discuss the different categories in terms of content, readability, and interactivity.

---

[9] http://reactable.com/

### III.1.1  Map content

Maps are used to display spatial content that is made of points, lines and areas, with accompanying text. All the different types of Accessible Interactive Maps that we previously described do not provide the same opportunities to render these different elements. DIMs that are explored with a tangible artifact mainly rely on auditory feedback. It is therefore possible to render maps that are made of several landmarks or separated areas such as states, but the type, number and resolution of rendered elements is obviously limited. Even though 3D sound may increase the quantity and quality of auditory feedback, maps from this category are quite simple in general. In fact, most of the prototypes based on 2D pointing devices were used to display choropleth maps, i.e. maps with regions only (Brittell *et al.*, 2013; Delogu *et al.*, 2010; Zhao *et al.*, 2008), or to display a very few number of landmarks (Milne *et al.*, 2011; Pielot, Poppinga, Heuten, & Boll, 2011).

These limitations are very similar when the DIM is directly explored with the fingers. Finger-based exploration prevents from having reference frames issues (when the visually impaired user does not exactly know where the device is pointing) but cannot provide more complex maps. In particular, it has been observed that the simple exploration of lines is problematic (Klatzky *et al.*, 2014). For example, Poppinga *et al.* (2011) reported that participants were not able to know whether two roads were close to each other or were crossing each other. They also had difficulties to identify the direction of short roads, and were less accurate when redrawing a "zoom-out" map than a "zoom-in" map, i.e. when more roads were displayed on the screen. Additional feedback can be provided with the vibrations of the devices, but the vibrations are not spatialized (i.e. the whole device vibrates), and hence cannot provide accurate cutaneous feedback. Consequently, most of the existing Digital Interactive Maps based on finger exploration are quite simple, and only display a very limited number of elements (Kane et al., 2011; Pielot, Poppinga, & Boll, 2010; Simonnet, Bothorel, Maximiano, & Thepaut, 2012; Su et al., 2010; Yairi, Azuma, & Takano, 2009).

DIMs that are based on less conventional tactile feedback (force-feedback, latero-tactile displays, Braille mice, etc.) can be more complex or detailed. For instance campus maps, alongside buildings plans, country maps with various cities and/or areas, street maps, etc., have been designed. Indeed, a number of cutaneous and haptic feedbacks can be used to render various elements such as boundaries, textures, points of interest, etc. (see Golledge *et al.*, 2005). Furthermore, haptic devices can help to recognize geometric properties of objects (Brewster & Brown, 2004).

Hybrid Interactive Maps are very useful for visually impaired users because the relief or the objects used to display the map represent useful tactile cues. As we already mentioned, Raised-Line Maps have been criticized for low resolution when compared to visual maps but they remain the best way for displaying complex content, using different patterns of points (e.g. triangles and circles), lines (e.g. dotted lines and plain lines), and areas (filled and half-filled). When making them interactive, Braille labels can be removed, and new elements can be added (Brock *et al.* 2015). Besides, the fact that they support two-handed exploration is highly beneficial (Wijntjes, van Lienen, Verstijnen, & Kappers, 2008a).

When compared to Interactive Tactile Maps, Tangible Maps are more limited by the number and type of elements that they can render. Indeed, physical icons are generally relatively large, which limits the number of points of interest that can be simultaneously represented (see e.g. Ducasse *et al.*, 2016). Only two prototypes make it possible to represent lines using either physical bricks

(Schneider & Strothotte, 2000) or retractable strings (Ducasse *et al.*, 2016). Although the former prototype allowed the construction of routes only, the latter allowed the construction of complex representations. There are many advantages to physical lines: users can easily locate and identify them, but also interact with them, which can be used to provide useful pieces of information on borders, rivers, transportation routes, etc. Yet, there is still a challenge on rendering areas with Tangible Maps.

Finally, Raised-pin displays are really adapted for rendering various patterns of symbols (Limin Zeng & Weber, 2012a) and lines (Ivanchev *et al.*, 2014), but their resolution is drastically limited by the size of the device and the number of pins. Then, even though it is possible to display various points of interests, lines and areas, the current prototypes cannot be used to display complex maps. Besides, several pins are needed to distinguish different symbols (Limin Zeng & Weber, 2012a), which requires a lot of space, and hence impacts map resolution. In fact, they cannot yet reach the complexity of Raised-Line Maps.

### III.1.2 Map readability

#### *Number of points of contact*

As we already mentioned, constraining the exploration of a map to a single point of contact can lead to cognitive issues. Indeed, in order to build a mental representation of the map being explored, users must perform a complex cognitive integration along space and time. They must integrate the path of their own movement and of the movement of the cursor with a transfer function if a pointing device is used. In addition, users must integrate the tactile cues under their fingers, which are related to each position over time (Klatzky *et al.*, 2014). This complex integration process is cognitively demanding, and may impact user's performances. For example, Loomis *et al.* (1991) showed that exploring a raised-line drawing with one finger only (i.e. one point of contact) is similar to exploring a visual drawing with a narrow field of view (the size of a fingertip). Not surprisingly, two-handed exploration of tactile images proved to be more efficient than exploration with one hand only (Wijntjes *et al.*, 2008a).

Among DIM prototypes, only the GRAB (Iglesias *et al.*, 2004) prototype, which relied on a force-feedback device with two handles, provided more than one point of contact. The evaluation showed that using a second point of contact helped users "orientate themselves in space, more readily understand object's relationships (distribution and distance) and makes re-finding objects easier".

Because they rely on a physical representation of the map, HIMs allow two-handed exploration. In addition, most of them provide multiple interactive points of contacts. Among Interactive Tactile Maps, only the earlier (see Abaplan for instance) relied on a single interactive contact point. More recent Interactive Tactile Maps (see e.g. Brock *et al.* 2015) provide many interactive contact points, which support multiple fingers and gesture command menus. Interestingly, Brock *et al.* (2015) observed that visually impaired users prefer not to activate any interactive feedback when they explore a map for the first time. Finger tracking in Tangible Maps can either rely on cameras (Schneider & Strothotte, 2000) or IR tracking (Ducasse, Macé, & Jouffrais, 2015). Even though in the prototype of Schneider *et al.* only one finger was tracked, it is now possible to track multiple fingers using a camera (for example using the CCV library). Then it would be interesting to design gestural interactions and evaluate their usability. As for raised-pin displays, they can detect touches and therefore offer a wide range of possibilities to in interact with the map.

### *Haptic frame of reference*

Another important aspect is the possibility offered to the user to build a reliable haptic reference frame during tactile exploration. We already mentioned than pressing a key or interacting with a joystick does not provide an external reference frame. These devices are referred as "isometric" devices, in contrast to "isotonic". Isotonic devices (also referred as "free moving devices") involve movements of the hand, and therefore provide more kinesthetic feedback than isometric devices, which is useful to build an egocentred reference frame. Zhai *et al.* showed that isotonic devices outperform isometric devices in position control tasks. However, Millar *et al.* (Millar et al-Attar, 2004) showed that using an external (allocentred) reference frame to encode spatial information resulted in better performances.

DIMs based on isometric devices (keyboard, joystick) do not provide a reliable external haptic reference frame. Indeed, the displacement of the cursor is relative to the previous position, and its current position within the map cannot be inferred by touching back the pointing device. Both Zhao *et al.* (2008) and Delogu *et al.* (2010) reported that using a keyboard made it difficult for the user to know the cursor position within the map. DIMs based on isotonic devices (mouse, handle) do provide a better feedback in order to build a haptic reference frame, but can still generate confusions if the cursor and the pointing devices are not always solidary. DIMs based on direct exploration provide a more reliable reference frame because the mapping between the hand and the map is fixed. In addition the outline of the touch-enabled device can be perceived, which represents efficient tactile landmarks. When using a camera to track the hand instead of a touch-enabled device, these tactile cues are missing. Then, some prototypes included a rigid frame that delimited the exploration area (see Bardot *et al.*, 2016).

Most of force-feedback devices also provide an external reference frame because the displacement of the device is in general constrained. The device cannot move over physical limits in space. Rice *et al.* (2005) reported that such a frame was very helpful. However, the physical limits of the device can be used as a reference frame if they correspond to a static view of the map. If the map view is displaced when the user pushes towards one side (sometimes called "inertial displacement"), then, in absence of efficient feedback, the reference frame is lost.

Obviously, HIMs provide a very stable and reliable haptic reference frame. Indeed, any static point of the tactile display (relief, identified item, edge of the display, etc.) can serve as a reference point, and anchor the reference frame. The reference frame can then be fed by all the other static tactile elements displayed in the map.

### III.1.3  Map interactivity

### *Updating map content, zooming and panning*

All the accessible interactive map prototypes that we described share a basic feature: the user can select a map element to retrieve its name. However, very few prototypes provided additional interactions with the map display (e.g. panning, zooming, filtering, highlighting) or eventually the map content (e.g. annotation, edition).

DIMs can be readily updated and are not limited by physical constraints. In that sense, they are similar to visual maps that sighted users can access online, and share the potential of providing visually impaired users access to a large quantity of geospatial data. However, performing zooming and panning operations on DIM, without any tactile cue, leads to sensory and cognitive challenges that have not yet been addressed in any experimental study.

Unlike DIMs, Hybrid Interactive Maps (HIMs) are shaped and constrained by the physical display. Interactive Tactile Maps are constrained by the tactile overlay, which cannot be dynamically altered. Different map contents, but also different views or different scales, must be rendered with different tactile overlays. Of course, it is possible to pre-print these different overlays, and dynamically call the corresponding digital content when one overlay is being used. However, when zooming or panning, users must interrupt the ongoing exploration in order to replace the overlay, and start a new exploration process after the corresponding digital content has been called. In order to link the mental representations corresponding to both maps, they have to find reference points that were on both overlays. This procedure clearly leads to cognitive challenges too. Furthermore, users cannot select a scale or a view that has not been previously prepared. At the intersection between Interactive Tactile Maps and Refreshable Maps, LineSpace (Erasable Tactile Map) does not provide regular panning and zooming operations. Instead, a new map with a different view or scale is printed over a blank space around the map being explored. However, the cognitive issues that we already mentioned (interruption of the current exploration and finding common reference points) also apply in that case.

The dynamicity of Tangible Maps is also constrained. Moving, adding or removing tangible objects is possible. Obviously, rescaling and repositioning take some time, but it renders the representation more flexible than regular tactile maps such as raised-line maps or 3D small scale models. Nevertheless, further studies are required to investigate whether Tangible Maps raise perceptual and cognitive issues. In the GeoSpace tangible and visual map (Brygg Ullmer & Ishii, 1997) users could rotate and rescale the map by moving two objects, but only a few elements of the maps were rendered using physical objects. Shaer *et al.* (2009) refer to this problem as a problem of *scalability:* when one user needs to zoom in or zoom out, all the objects need to be repositioned, i.e. the whole map needs to be reconstructed.

Refreshable Tactile Maps are the most dynamic interactive maps. It is possible to update the content instantly, but also to provide advanced interactive functions such as zooming and panning operations (Shimada *et al.*, 2010) or annotation (Limin Zeng & Weber, 2012b).

### *Adapted exploration functions*

Additional functions that help visually impaired users to explore maps are not specific to DIMs or HIMs. They can provide verbal descriptions and guidance, spatial and semantic filtering, but also specific computations (e.g. distance between two points). Some functions have been implemented and evaluated.

As we previously mentioned, it is a challenge to find and relate specific points when the exploration is tactile, especially when it is performed with only one contact point. For instance, Kane *et al.* (2011) developed three interaction techniques to help the user locate, relocate and relate points of interest on a map. The Talking TMAP prototype (Miele et al., 2006) provided assistance to find a location, and calculate distances, but also provided a menu for modifying the settings (sensitivity, unity of measure, and speech rate).

With the Interactive Raised-Line Map called Mappie (Brule *et al.*, 2016), children could also rely on audio instructions to locate specific points of interests. In addition, they were able to choose among different types of spatial content (city, countries, etc.) for the same overlay, which is very useful for teachers. Indeed, the same tactile overlay may be used in conjunction with different digital contents. It may then serve successive lessons of the same discipline, but also different disciplines. A tactile overlay representing a country can for instance accompany a progression in Geography when digital content is progressively uncovered. The same overlay can also serve a lesson in Economics with a different digital content.

Other prototypes provided access to complex functionalities. For instance, iSonic (Zhao *et al.*, 2008) included a "gist" of the map, which was a sequence of sounds that provided users with an "overview" of the regions. A subpart of the map could then be selected. The iSonic prototype provided additional functions such as: details-on-demand, adjust information level, situate (give the current status of the interface), select (only the regions selected triggered audio feedback and are played for the gist), brush (between the table and the map views), filter and search. Bardot *et al.* (Bardot et al., 2014, 2016) also implemented several interaction techniques for spatial or semantic filtering of the map content. A first one allowed the user to get a spatial overview or spatial filtering of the map before being guided towards a specific point of interest. A second one enabled users to select the content to be displayed, and then avoid the cumbersome exploration of undesired items.

## III.2 Geographic maps for visually impaired people and spatial cognition

Maps serve a concrete purpose: acquiring spatial knowledge about an environment. In the following sections, we present studies on acquisition of spatial knowledge without sight that have been done using tactile maps or interactive maps.

### III.2.1 Tactile maps and spatial cognition

Some studies specifically investigated the benefits of tactile maps for spatial learning in visually impaired adults. Jacobson (1992) compared sketch maps drawn before and after reading tactile maps. He observed that the sketch maps drawn after map reading included far more description and details than the initial sketch maps. In another study (Jacobson & Kitchin, 1995), visually impaired adults failed in estimating relative distances between towns, but succeeded in locating towns on a partially complete tactile map. Furthermore they successfully determined which map was correct out of a set of three rotated maps. In a follow-up study (Jacobson, 1998b), sketch maps proved to be more accurate for a group of participants which had explored an audio-tactile map compared to subjects having walked the same route with a mobility instructor. Espinosa *et al.* (1998) observed that participants who learnt a route from a combination of direct experience and tactile map reading had a better spatial knowledge of the environment than those learning the route from direct experience only or verbal descriptions. In a second experiment, participants performed just as well after exploring a tactile map as after direct experience in the environment. Similarly, Caddeo *et al.* (2006) observed that participants who had access to a tactile map showed better performance in walking time and a reduced deviation from the route as compared to participants directly experiencing the environment with or without verbal descriptions. Similar results have been found, when studying tactile maps with visually impaired children (Ungar, Blades, Spencer, & Morsley, 1994).

Different hypothesis explain why tactile maps are well adapted for acquiring spatial knowledge in the absence of vision. Tactile maps preserve relations between landmarks in space but present those relationships within one or two hand-spans (Ungar, 2000). Thinus-Blanc and Gaunet (1997) argue therefore that exploration of a tactile map demands a smaller working load than exploration of a real environment. Also they outline that during tactile exploration of space, it is possible to keep a fixed reference point, whereas during the exploration of space via locomotion, the participant is moving and so is the reference system (the own body). Besides, the tactile map is simplified in content and therefore free from the perturbations that can be present in the environment (Ungar, 2000).

### III.2.2 Interactive maps and spatial cognition

Several studies investigated if interactive maps can be used for acquiring spatial knowledge by visually impaired people. Very often these studies compared different groups of users including early and late blind, but also people with residual visual perceptions, and blindfolded sighted.

First we report the studies that have been done with blindfolded sighted people, and which may present limited validity due to the perceptual and cognitive differences between visually impaired and sighted people. Poppinga *et al.* (2011) asked eight sighted users to explore a smartphone application with vibration and audio output and draw a map of the perceived environment. They compared two zoom levels and observed that the "zoom in" condition resulted in a more accurate drawing than "zoom out" condition. Participants correctly perceived basic information concerning the map, but also relations between map elements. However, the authors suggested that the task was cognitively demanding as participants needed up to 15 minutes for sketching a rather small map. In another study (Lohmann & Habel, 2012), twenty four blindfolded sighted subjects compared two conditions of a DIM prototype with speech output. In the first condition only names of landmarks and routes were indicated. In the second condition, additional information about the relationships between geographic items was provided. Participants acquired more precise spatial knowledge in the second condition. However, the result also depended on the type of spatial knowledge. Scores related to landmarks showed a larger difference between the two conditions than scores related to routes. Pielot *et al.* (2007) evaluated a DIM with a tangible pointing device with eight blindfolded sighted participants. The results showed that the tangible pointing device, called a virtual listener, allowed detecting small deviations from the real orientation. Milne *et al.* (2011) compared two prototypes of a DIM with five blindfolded sighted participants. One device was based on a stylus while the other was based on body-rotation. They observed issues related to the shifting of reference frames between egocentred and allocentred perspectives.

Other studies were done with blind participants and therefore present a higher validity. Among these studies, a few evaluated DIMs based on touchscreen devices and audio output. Jacobson (1998) asked five visually impaired and five blind people to evaluate a map prototype based on a touchpad with auditory feedback. He analyzed verbal descriptions, map drawings, and qualitative feedback. He observed that all users successfully created mental representations following the DIM exploration. Besides, the users found the interface simple, satisfying and fun. In the study of Heuten *et al.* (2007), eleven blind users explored a DIM based on a touchpad and a stylus, with the instruction of understanding spatial relationships between map elements. Users found the exploration easy, except for the identification of the shapes of the elements. The authors also observed confusions when two similar objects were close to each other. Yairi *et al.* (2008) asked four blind people to explore a DIM with musical feedback, and then walk the route unaided. All participants reached the goal, even if one was unsure about it. Even if people made wrong turns at cross points or felt lost, they were able to correct their route. Simonnet *et al.* (2009) observed two blind sailors learning a maritime environment with a haptic device and then navigate at sea. Their study revealed that using the map was beneficial because the users had to mentally coordinate egocentred and allocentred maps. In a more recent study, Picinali *et al.* (2014) compared the results of five blind participants who walked along a corridor versus five blind participants who explored the DIM with a joystick. The results showed that all participants were able to build correct mental representations that were similar to the reference map. All these studies show that DIMs with audio output can effectively be used for creating mental maps.

As we mentioned, DIMs can be augmented with vibrations. In the study of Simonnet *et al.* (2012), one blind user explored a DIM based on a tablet with auditory and vibratory feedback. After exploration, the user had to draw a map of the explored environment. They observed that the drawing was relatively precise. In the study of Yatani *et al.* (2012), ten blind and two low-vision participants explored a DIM based on a smartphone with audio and vibratory output, but in two different conditions. In the first condition, the subjects used the smartphone with auditory output only. They received additional vibratory feedback through nine motors in the second condition. They were then asked to draw sketch maps. The drawings were more accurate with additional tactile feedback.

A few studies investigated the acquisition of spatial knowledge with HIMs. Jacobson (1996) compared route learning with a mobility instructor versus with an Interactive Tactile Map (touchscreen with tactile overlay). The author used many methods including verbal description, map drawing, distances by ratio-scaling, tactile scanning assessment, and talk aloud protocol. The results showed that both groups were able to complete the route and verbally describe it. All sketches showed a high degree of completion and correctness but the group who had explored the interactive map was more accurate. Participants also provided positive qualitative feedback. More recently, in a study with twenty-four blind participants, Brock *et al.* (2015) compared the usability of an Interactive Tactile Map versus a regular raised-line map with braille legend. They measured the time required to explore the maps, the correctness of the spatial knowledge acquired during exploration, and the satisfaction. The results showed that Interactive Tactile Map was significantly more efficient, and preferred among users.

Finally, Zeng and Weber (2014) evaluated the acquisition of spatial knowledge using a Refreshable Map with ten blind users. They compared it to two other conditions including a regular raised-line map, and a DIM displayed on a tablet. They measured the subjects' performances in reading street names and preparing a journey. The results showed that participants were able to perform these tasks using the raised-pin device or the raised-line map, but failed to perform them on the tablet.

In conclusion, it appears that DIMs can be used to acquire spatial knowledge but, except in the study of Yairi *et al.* (2008), spatial learning was limited to simple topological features (relative positions of items within the map).

In some of these studies, but also in the literature in psychology, there have been contradictory results showing that early blind, late blind, visually impaired, and sighted subjects outperform each other (Cattaneo & Vecchi, 2011). Other factors, including expertise in tactile reading, education level, but also the type of drawing being read, might have been confounded with the visual status. In any case, we suggest that a greater access to interactive maps is mandatory for visually impaired users and will decrease the differences between the different groups of subjects.

### III.3 Maps and other graphics

The challenges related to the accessibility of maps for visually impaired people are similar to those observed when designing accessible interfaces for other types of interactive graphics. By graphics, we refer to a variety of materials whose layout is used to provide spatial content to the reader including diagrams, figures, drawings, as well as maps (The Braille Authority of North America, 2010). In this section, we present a non-exhaustive list of Interactive Graphics prototypes. They illustrate how interactive devices providing non-visual access to maps and graphics are similar. As for maps, it is easy to classify the Interactive Graphics prototypes in two categories (Digital and Hybrid Interactive Graphics, which we call DIGs and HIGs) and their subcategories.

#### III.3.1 Examples of interactive graphics

A number of DIG prototypes have been based on touch-sensitive surfaces or motion capture. For example, the AudioFunctions prototype combines a novel sonification technique with touch interaction to enable a visually impaired to explore a mathematical function (Taibbi, Bernareggi, Gerino, Ahmetovic, & Mascetti, 2014). Users performed better using the prototype than using a raised-line diagram. Other prototypes (Gerino, Picinali, Bernareggi, Alabastro, & Mascetti, 2015; Yoshida, Kitani, Koike, Belongie, & Schlei, 2011) enable visually impaired users to identify simple shapes displayed on a touch-screen device. TouchMelody (Ramloll & Brewster, 2002) augments raised-line diagrams with spatial non-speech sounds. Both the index of the non-dominant hand (used as a reference point) and the index of the dominant hand (used to explore the diagram) were tracked by a motion-capture system, and a sound was played according to the position of one index as compared to the other one.

Force-feedback devices have also been used. The prototype designed by Yu and Brewster (2003) could be used with either a Geomagic Touch X device or a Logitech WingMan Force Feedback mouse, and enabled visually impaired users to explore line graphs or bar charts. McGookin and Brewster (2007) designed a prototype that enabled users to drag the bar of a bar chart. Bernareggi *et al.* (2008) developed a system that enabled users to insert, delete, connect or disconnect nodes.

Giudice *et al.* (2012) combined gestural input with audio and vibratory feedbacks. When users moved a finger over the tablet, vibratory patterns indicated whether they were touching edges or vertices. This prototype proved efficient for the exploration and understanding of bar charts as well as for letter recognition tasks and orientation discrimination tasks.

Pointers with additional tactile feedback have been used in other prototypes. Wall and Brewster (2006) used a stylus for pointing combined with a mice with an array of pins (VTPlayer). The user pointed to different zones of the graphs, and received tactile cues according to the section of the pie chart being explored. Pietrzak *et al.* (2007) used a similar device to provide directional cues that guide the users during the exploration of geometrical shapes. Levesque *et al.* (2008) showed that simple and small shapes could be rendered using the STRESS latero-tactile display and three primitive drawings (dots, vibration and gratings).

Examples of HIGs include the tangible prototype for the non-visual exploration of graphs by McGookin *et al.* (2010). This system combined a fixed grid and movable objects that represent the top of a bar or the turning point of a linear function. When moving a slider along the x-axis, the user was able to retrieve the corresponding y-values, which were sonified. Similarly, TIMMs (Manshad, Pontelli, & Manshad, 2012) are objects that provide multimodal feedback and enable blind persons to create and modify graphs and diagrams.

### III.3.2 Maps and graphics: a comparison

As previously mentioned, raised-line maps are bulky, not interactive, and their accessibility can be improved by making them interactive and dynamic. The same applies to any type of raised-line graphics. So far, most research projects have either focused on graphics or on maps. Obviously, researchers could benefit from taking into account both of these application areas. Yet, it remains to be explored to what extent the issues raised by interactive maps are similar to those raised by interactive graphics.

One essential point is that all graphics, whether they are drawings, maps, bar charts, diagrams, etc., are solely made of four primitives: dots, lines, areas, and labels. Moreover, colors and textures are often used in order to improve readability. Graf (2010) distinguishes the propositional representation (verbal annotations) from the spatial representation (the map itself that represents the topology of the environment). Considering multiple dimensions (including layout, complexity, dynamicity and usages), we did not find any significant difference between the properties of these two representations related either to diagrams or maps. Therefore no matter the field of application, findings concerning the legibility of the symbols and representations with a particular content or display can apply to any other content or display. For example, above we cited a few articles that aim at understanding how simple geometrical drawings that are displayed on a touch-screen can be identified using audio and vibrational feedbacks (e.g. Giudice *et al.*, 2012; Lévesque & Hayward, 2008). These findings could inform the design of both interactive maps and graphics prototypes. Obviously, we would need to perform comparative experiments in order to ensure that, except slight differences in complexity, there are no specific needs according to the type of graphics that is rendered.

## III.4 Ongoing and future research in interactive tactile graphics

There are a few topics that we wanted to address which represent current challenges for research in accessible interactive graphics.

### III.4.1 Rich open and volunteered data

In this chapter we have mainly discussed the design of accessible interactive map prototypes. Yet, for the usability of maps, the availability and reliability of geographic content is as important as the design of the devices. Indeed map prototypes will not be used outside of research laboratories if adapted geographic data are not available. In Section II.2.4, we presented a few projects that aim at automating the production of adapted content or at facilitating the collection of volunteered geographic data. We suggest that OpenStreetMap is particularly relevant as specific accessibility tags such as tactile paving can be added, (El-Safty, Schmitz, & Ertl, 2014). Then, visually impaired users, but also sighted users that want to participate, can create annotations when they are travelling or exploring the map (Holone & Misund, 2007; Rice *et al.*, 2013). Sighted users can also volunteer to add details based on street view images (Hara *et al.*, 2013) or on their own knowledge of the places.

### III.4.2 Authoring tools and content sharing

Rich and adapted open data is not the only challenge. Up to now, the automatic production of maps is still difficult and the intervention of professionals is required (see I.1). Besides, authoring tools are not common and that they are mainly circumscribed to research projects. Researchers should closely work with tactile graphic specialists in order to better understand how the production of adapted content could be further automated. Authoring tools may then include adaptation functions that help experts but also non experts to create accessible content that is adapted to be displayed on DIMs or HIMs. Such tools may encourage professionals to create and share accessible map content. Successful projects have been developed (e.g. Götzelmann & Pavkovic, 2014b; Kaklanis et al., 2011, 2013a) and should now be tested in-the-field, to evaluate their long-term impact upon the accessibility of graphical data for visually impaired users.

### III.4.3 Shape-changing interfaces

Interactive Tactile Maps present a high usability because they provide reliable tactile cues and haptic reference frame, as well as interactivity. However, they are constrained by the physical display, which prevents dynamic zooming and panning. In contrast, refreshable displays, which include both physical and digital representations, offer remarkable possibilities for dynamic interaction with maps and graphics (including, zoom, pan, annotation, filtering, etc.). In addition, they enable users to explore the maps with both hands, which is more efficient. However, those devices are currently very expensive, which prevents a large adoption by visually impaired users and professionals. Moreover, current prototypes only provide a small surface size.

Low-cost and large refreshable displays are still at infancy, but a number of approaches are promising. Shape memory alloy actuators change of shape when they are heated up. Voice-coil motors and piezoelectric actuators may also provide solutions for larger displays (see Vidal-Verdú & Hafez, 2007; O'Modhrain et al., 2015 for reviews). Wilhelm et al. (2014) developed a prototype based on microfluidic phase change actuators. The actuators are filled in with a phase change material that can be heated. When pressure is applied, the membrane laid over the display is bulged. Taher et al. (2015) investigated new interactions with physical bar charts. Other research works have to be mentioned, such as Relief (Leithinger & Ishii, 2010), inFORM (Follmer, Leithinger, Olwal, Hogge, & Ishii, 2013), and Lumen (Poupyrev, Nashida, Maruyama, Rekimoto, & Yamaji, 2004). These physical visualization devices could increase the accessibility of graphics for visually impaired users in the future.

### III.4.4 3D printing with embedded interactivity

There are several limitations to 3D printing such as the relatively high cost for acquiring a printer, time for printing, limited size of the printed object, and lack of interactivity once printed. However, it is an emerging tool for the production of maps, graphics and books for visually impaired users. We already described the work of Götzelmann and Winkler (2015) who automated the production of interactive 3D maps, which in contrast to raised-line maps can provide multiple levels of relief. Many other recent studies show the importance of 3D printing for visually impaired users.

Buehler et al. (2014) reported that the creation of educational materials is one of the three primary functions of 3D printing in special education. With 3D printing, it is possible to print customized objects that can be used to explore or annotate maps. For example, users can explore a virtual map by moving and rotating a toy above a table (Pielot et al., 2007). Such a toy could be personalized using 3D printing in order to get the students more engaged with the exploration task. Brulé et al.

(2016) reached the same conclusion. They used 3D printing to create tangible objects that can be used in association with an Interactive Tactile Map device. Such objects could be placed on a tactile map to augment the information that was displayed or to highlight important elements. Giraud and Jouffrais (2016) showed that 3D printing with low cost prototyping resources empowers specialized teachers to create their own teaching material. Swaminathan *et al.* (2016) proposed a sense-making platform for blind people using dynamic 3D printing. Finally, Gual, Puyuelo, and Lloveras (2014) compared the use of 3D printed symbols versus 2D tactile symbols in a tactile map. They found that 3D symbols were easier to memorize than 2D symbols.

Recent work has also shown that 3D printed objects are not limited to non-interactive plastic structures. Current 3D printers can print soft and interactive objects that embed conductive filaments (see e.g. Peng, Mankoff, Hudson, & McCann, 2015). Objects produced with those printers could greatly enhance the interactivity of Tactile Maps and Graphics as they could potentially vibrate, move, emit sounds, or detect how they are grabbed.

## IV. Conclusion

In this chapter we presented an overview of accessible interactive maps for visually impaired people. We identified two families of Interactive Maps that differ according to the presence or absence of a physical representation of the map, which is useful for visually impaired users because it is perceivable by touch. The first family was called Digital Interactive Maps (DIMs) and relies on a digital representation of the map only. The second family was called Hybrid interactive Maps (HIMs) and relies on both digital and physical representations of the map. We defined subcategories in each family that, hopefully, may help to structure the research field.

In each family, we have observed a large variety of prototypes based on various input and output interaction devices. They have leveraged the design of non-visual interaction techniques allowing visually impaired users to explore a map. Additional functions have been designed too, allowing zooming, panning, annotating, sharing, visualizing over time, etc.

However, these different devices come with advantages and shortcomings. DIMs based on pointing devices such as touch-enabled screens or video tracking are available at low prices and thus affordable for everyone. They can easily be used in many situations (home, school, mobility, etc.) but they miss tactile cues that are useful for non-visual tactile exploration because they facilitate tactile integration, and also provide a reliable haptic reference frame. Hence they must be designed for specific conditions such as mobility, and for compensating the absence of tactile cues. In addition, spatial content should not be too complex.

Based on a physical display, HIMs are more adapted to non-visual exploration. We observed that one type of HIM—Interactive Raised-Line Maps—has been largely addressed in the literature, both on design and evaluation sides. Because they have a high usability, they are now spreading in the wild, and devices are being used in specialized centers. Of course, they still suffer from the necessity of printing raised-line maps in advance, which is manageable but costs time and money. Refreshable displays supposedly provide an alternative; however, they are not yet available with a sufficient resolution at affordable cost. Tangible Maps have not been extensively studied so far. With the spreading of TUIs, but also 3D printers and low-cost prototyping technologies, we guess that, during the upcoming years, these devices will get further addressed as a research question, but also more used in the wild. Indeed, although spatial resolution will always be limited by the size of the physical icons, they provide an appropriate setup to explore and manipulate spatial data without vision, especially in collaboration with other sighted or non-sighted users.

We have to mention that, nowadays, it is still difficult for visually impaired people to access geographic information. Yet, non-visual access to geographic information is crucial for education, as well as mobility and orientation. It has significant consequences on personal and professional occupations, and on social participation. Therefore, it is very important that future research work on maps focuses on the design and evaluation of interaction techniques that are prospective, but also on devices that are readily usable. As an agenda for short- and mid-term research, we suggest topics that should be addressed: improving the accessibility of Digital Maps with wearable technologies; facilitating the autonomous (without the intervention of a tactile document maker) making of Interactive Maps; and designing interaction techniques that provide visually impaired people with more interactive functions such as zooming, panning, annotating, and collaborating. Obviously, researchers and designers should always keep in mind that maps serve a purpose: the acquisition of spatial knowledge. In order to validate that the devices effectively serve this purpose it is necessary to conduct user studies with visually impaired users. When possible, new displays should be studied in-situ, i.e. outside the lab, to better understand how and why interactive maps and graphics are used.